\documentclass[a4paper,fleqn]{cas-dc}

\usepackage[authoryear]{natbib}
\usepackage{todonotes}

\def\tsc#1{\csdef{#1}{\textsc{\lowercase{#1}}\xspace}}
\tsc{WGM}
\tsc{QE}

\ExplSyntaxOn
\keys_set:nn { stm / mktitle } { nologo }
\ExplSyntaxOff

\ExplSyntaxOn
\RenewDocumentCommand \eadauthor {} 
    { 
      \seq_map_inline:Nn \l_stm_au_seq 
        { 
            \regex_extract_once:nnNTF {(\w)\w*-(\w)} { ##1 } \l_stm_au_fn_seq
            { 
                \seq_pop_left:NN \l_stm_au_fn_seq \temp_var
                \seq_use:Nn \l_stm_au_fn_seq { .- }
                { . } 
            }
            { 
                \regex_match:nnTF { \. } { ##1 } 
                { ##1 }
                { \tl_head:n {##1}. }
            }
      }{ ~\l_stm_au_sn_seq }
    }
\ExplSyntaxOff

\begin{document}
\let\WriteBookmarks\relax
\def\floatpagepagefraction{1}
\def\textpagefraction{.001}

\shorttitle{I'm stuck! Debugging computational solid mechanics models}   

\shortauthors{Comellas, Pelteret, and Bangerth}  

\title [mode = title]{I'm stuck! How to efficiently debug computational solid mechanics models so you can enjoy the beauty of simulations}

\author[1]{Ester Comellas}[orcid=0000-0002-3981-2634]
\cormark[1]
\ead{ester.comellas@upc.edu}
\affiliation[1]{organization={Serra Húnter Fellow, Department of Physics, Universitat Polit\`{e}cnica de Catalunya},
            addressline={Rambla Sant Nebridi, 22}, 
            city={Terrassa (Barcelona)},
            citysep={}, 
            postcode={08222}, 
            country={Spain}}

\author[2]{Jean-Paul Pelteret}[] 
\ead{jppelteret@gmail.com}
\affiliation[2]{organization={Independent researcher},
            country={Germany}}

\author[3]{Wolfgang Bangerth}[orcid=0000-0003-2311-9402]
\ead{bangerth@colostate.edu}
\affiliation[3]{organization={Colorado State University},
            addressline={Department of Mathematics, Department of Geosciences}, 
            city={Fort Collins},
            postcode={80524}, 
            state={Colorado},
            country={United States}}


\begin{abstract}
A substantial fraction of the time that computational modellers dedicate to developing their models is actually spent trouble-shooting and debugging their code. 
However, how this process unfolds is seldom spoken about, maybe because it is hard to articulate as it relies mostly on the mental catalogues we have built with the experience of past failures.  
To help newcomers to the field of material modelling, here we attempt to fill this gap and provide a perspective on how to identify and fix mistakes in computational solid mechanics models. 

To this aim, we describe the components that make up such a model and then identify possible sources of errors. 
In practice, finding mistakes is often better done by considering the \textit{symptoms} of what is going wrong. As a consequence, we provide strategies to narrow down where in the model the problem may be, based on observation and a catalogue of frequent causes of observed errors. In a final section, we also discuss how one-time bug-free models can be kept bug-free in view of the fact that computational models are typically under continual development.

We hope that this collection of approaches and suggestions serves as a ``road map'' to find and fix mistakes in computational models, and more importantly, keep the problems solved so that modellers can enjoy the beauty of material modelling and simulation. 
\end{abstract}


\begin{keywords}
computational solid mechanics \sep 
material modelling \sep 
constitutive model \sep
finite element method \sep
debugging \sep
numerical implementation \sep
numerical algorithm \sep
\end{keywords}

\maketitle

\section{Introduction}
\label{sec:intro}
Scientists often live for those few, short moments where everything comes together in one table or graph that contains the fruits of weeks of work. The time between these moments is typically filled with building things (experiments, measurement devices, software) and endless sessions of trouble-shooting and debugging. 

Yet, we rarely talk about these often frustrating periods in the lives of scientists, or the lessons we could learn from them. In part, this is because finding the causes of why what we built is not working is as much an art as it is a science: A lot of it relies on mental models, experience we have built from past failures, and recall of mistakes we have made before and what helped us find them. Nearly everyone asked, for example, to describe their process of debugging a piece of software draws a blank: We know how to do it, and some of us are good at it, but few of us can articulate how precisely we approach the task.

In this paper, we attempt to provide a perspective of how we go about deriving, implementing, testing, and debugging material models for computer simulations of mechanical objects -- the things that are all necessary to ultimately end up with that one graph or visualisation that shows that what we came up with matches what our experimental colleagues have observed!

\paragraph*{Computational solid mechanics.} The concrete case we will be considering herein is how to get computational mechanics models right -- and specifically what to do when, as seems to always be the case on first try, a model does not seem to be right. In our context, a computational model consists of a mathematical formulation of a mechanical object's behaviour, an algorithm to solve the problem (often using numerical approximations), and a software implementation of this algorithm.
Such computational models have become instrumental for technological advancement in many fields of science and engineering as they provide a cost-efficient, safe, and environmentally friendly tool to explore and improve designs, manufacturing processes, testing set-ups, and certification procedures in a wide range of applications. They also provide crucial checks on whether our understanding of a complex material is consistent with its actual experimental response. Because we keep coming up with new and more complex materials, there continues to be a need for the development of new and different models.

In order to restrict ourselves to a concrete set of computational models for which we can provide advice, let us specifically turn our attention to situations where material behaviour is formally characterised by a \textit{constitutive relation}. This relationship defines the response of the material (typically its deformation and stress state), to internal and/or external stimuli (usually the action of forces or applied external fields, e.g. electrical or magnetic). 

What constitutes a ``successful model'' may actually be debatable. For our purposes here,  modellers will generally agree that a good model must be built on appropriate assumptions and meet the purpose that motivated its development. In other words, the material model must be based on well-founded hypotheses and predict physical, measurable outcomes to within a reasonable accuracy. At the same time, good models are as simple as possible and do not rely on parameters that can not be physically motivated.

Conversely -- and the focus of this paper -- we consider a model ``not successful'' if its computational predictions are either not physically reasonable or at least do not match what we physically measure using its ``real-world'' equivalent. We will then say that the model has mistakes, bugs, or problems, and that we need to ``fix'' or ``debug'' it.

\paragraph*{Goals.}
Defining and implementing such ``successful models'' is a many-step process that involves not only (i) defining the constitutive relations, but then also (ii) deriving the partial differential equations that govern material response, (iii) posing appropriate boundary and initial conditions, (iv) implementing this mathematical model in software, (v) assessing the correctness of the software's output, and (vi) possibly additional steps. As mentioned above, a first go-around of these steps rarely leads to a correct end result: The program's output will either be obviously unphysical, or simply not predict physical measurements on the actual object reasonably accurately.
Herein we provide a framework for how to think through where in this list of steps the problem may lie. 

When talking about finding mistakes, it is often instructive to newcomers to a field to note that even long-timers spend more time fixing their mistakes than coming up with the first version. For example, even good programmers spend more time debugging their codes than writing them in the first place. As a consequence, we will highlight the mindset that implementing computational models is a challenge that more than anything else relies on \textit{experience} and on having a \textit{mental catalogue} of what typically goes wrong. 
Our ultimate goal is to provide ``road maps'' one can use to find mistakes in computational models.

\paragraph*{Non-goals.} When developing software, a substantial time is spent on finding coding errors that include compiler errors, segmentation faults, out-of-bounds accesses in arrays, dangling pointers, and similar things. These are real problems with computational models, but we will not address them here as they are not specific to computational mechanics -- good introductory books on programming 
will cover strategies to deal with these issues, the most important of which is to carefully \textit{read the error message}. Instead, we will for the most part assume that the simulation code for a computational model actually runs without error messages, but does not produce the expected output for reasons that are unrelated to simple programming mistakes.

\paragraph*{Intended audience.}
Seasoned professionals have often found useful ways to check their work. 
For example, \citet{Wilson2014, Wilson2017} provide a set of good practices for scientific computing to improve the productivity and reliability of the software developed.
But, as pointed out above, we rarely talk about ``debugging models'' in their entirety, and this contribution is an attempt at addressing this gap. 
Therefore, we intend this paper to be most useful to modellers starting as independent researchers, such as PhD or graduate students who already have a Masters degree.
We hope that it will also be useful for someone moving into the research system who is programming their own computational model for the first time.

In view of this target audience, this paper is a collection of approaches and perspectives that we, the authors, wish we had had when we started in our careers in material modelling.
That said, we believe the guidelines we provide herein are equally applicable to the broader computational modelling community.

\paragraph*{Outline.}
In Section~\ref{sec:model-outline} we will first come up with a concrete list of components that go into a computational model. Then, in Section~\ref{sec:what-is-wrong} we will summarise approaches on how to ``debug'' models that consist of these components, along with problems we have found often happen in each step.
In Section~\ref{sec:keeping-it-solved} we will provide strategies towards keeping working computational models ``bug-free'' as one continues to expand and build on them.
To conclude, Section~\ref{sec:conclusions} provides some final considerations on the trials and tribulations of debugging, and how it is an integral part of developing computational models.

\section{Components of computational solid mechanics model}
\label{sec:model-outline}
It is instructive to start by defining what exactly goes into a computational model of a solid mechanics system. This is because when a discrepancy arises between the model's output and what we expected the output to be (from physical considerations, or because we compare with actual measurements or a known analytical solution), it provides us with a road map of things we can individually examine. Conversely, if we have not considered something part of the model (say, certain types of boundary conditions) but we have implemented this piece wrongly or not at all, we may not consider it a source of the problem.

The following is a reasonably comprehensive set of steps one has to go through to define a complete computational model of a mechanical object:

\begin{enumerate}[1.]
    \item \textbf{Identify the purpose of the model.} Computational models are often developed to test scientific hypotheses, whether it is to elucidate underlying physical mechanisms, to perform \textit{in silico} experiments, to better characterise material behaviour, or to test new design ideas. Alternatively, they may be part of the validation and certification process in industrial designs. We must think carefully about the goal of our model, since it will dictate (or constrain) its theoretical framework and numerical implementation.

    \item \textbf{Establish the theoretical framework.} Starting from first principles, we derive the strong form of the governing equations that describe the solid mechanics problem at hand. Pairing the result of this process, which can be done entirely with pen and paper, with a constitutive relationship (material law) leads to a complete set of differential equations from which we then obtain the weak form of the governing equations by differentiation\footnote{%
    It is sometimes possible to use a variational setting to formulate the problem, e.g. if an energy functional can be defined \citep{Zienkiewicz2005a}. Differentiation of this functional then renders the weak form of the problem, equivalent to starting with the disparate strong form and constitutive law.%
    }.
          
    \item \textbf{Set up the pseudo-algorithm.} Often overlooked, this step requires translating the theoretical framework into a numerically-compatible description. To this aim, we must ideate a solution strategy. The continuous equations previously derived must be discretised in space (and possibly in time), adequate integration algorithms must be selected, and all necessary components, e.g. tangent stiffness matrix and right-hand side vector, should be identified and clearly defined. The result is a schematic of the complete algorithm to implement, which gives us a clear idea of the main elements in our problem and how they relate to each other. It is often useful to also come up with consistent notation that can then be used one-for-one in the computer implementation.
    
    \item \textbf{Define the numerical experiment.}
    On the basis of the guiding algorithm, we can devise multiple virtual experimental set-ups. Each experiment will require the definition of a specific geometry, boundary and loading conditions, and other model parameters which must be defined and implemented into the code.
    
    \item \textbf{Complete the numerical implementation.} We translate the pseudo-algorithm of our theoretical framework and the set-up of the numerical experiment(s) into ``real'' code. For this, we must choose a programming language and environment, as well as select adequate libraries and numerical tools~\citep[e.g.][]{MFEM,dealII94,Kratos,libMesh,FEBio,NGSolve,FenicsxBasix}. 
    Key aspects include, but are not limited to, programming paradigm (e.g., object-oriented), parallel computing functionalities~\citep[e.g.][]{mpi40,TBB}, as well as memory access and management. 
    Input and output interfaces must also be defined -- in particular, compatibility with visualisation tools ~\citep[e.g.][]{ParaView, VisIt, PyVista} -- in addition to other coding considerations like code extensibility.
    
    \item \textbf{Verify the model.} Once the computational model is set up and running, we must check that it accurately represents our conceptual description and specifications. In other words, has the pseudo-algorithm been correctly translated into the code? And, more importantly, are the code and its post-processing mechanisms doing what we expect them to do?
    
    \item \textbf{Validate the model.} Validation entails ensuring that the model is an accurate representation of the real world, within the context of its intended use. We typically use benchmark tests to compare results with those of similar codes, or with available experimental data.
\end{enumerate}

The result of all of these steps is a computational solid mechanics model, whose main parts are: 
(i) the definition of the geometry, model parameters, boundary conditions and initial conditions, including user input data;
(ii) the discretised governing equations, including the constitutive model that dictates the material behaviour;
(iii) the numerical algorithm used to solve the problem, whose core component is the solver; and
(iv) the output of results.
Once the computational model is set up and working, we are ready to use the code to explore ideas, advance the state of the art, answer our scientific queries, and produce that table or graph to visualise our findings!

\section{What could possibly go wrong?}
\label{sec:what-is-wrong}
Unfortunately, getting a computational model to work properly is not generally as easy as the previous section might suggest.
Whenever a model is not successful in the sense outlined in the introduction, it is important to recall that at least in principle, \textit{the problem may be with any of the steps listed in the previous section}. It is not useful to rule out some steps \textit{a priori} because it may lead to long debugging sessions of parts of the model that are not, in fact, wrong.

In order to stress the importance of keeping an open mind about finding where the bug may be, let us mention a tautological, but nevertheless useful, observation: When observing that a model is not successful, we typically (i)~assume that we have derived and implemented the model correctly (as is human nature), but (ii)~observe that the output is wrong. These two statements are in obvious conflict: They cannot both be true\footnote{There are situations where we believe that the output is wrong when in fact it is not. For example, we think that a particular physical set-up should yield a solution that is left-right symmetric, when that is not actually true and so observing a non-symmetric solution does not imply that the model is wrong. However, these cases are relatively rare and we will in the following assume that the model output is in fact wrong.}. 
Then it is worth noting that because \textit{our trust in the correct derivation and implementation of the model was apparently misplaced}, we ought to be suspicious about believing that certain parts of it are correct: \textit{A better approach is to assume that \textbf{any} component of the model is now suspicious and needs to be checked.} 

Of course, a model may consist of many pages of derivations and thousands or tens of thousands of lines of code. It is not productive to work through them top to bottom -- this would amount to trying to find the needle in the haystack by removing one hay stalk at a time until we have found the needle. We need a better strategy that helps us identify which general component might cause the issue in a first step, before we look at a smaller scale.

On account of these thoughts, let us below first outline some general considerations about narrowing down which component a problem might be located in, followed by discussions about typical problems in each of the components listed in the previous section.

In practice, ``there is always one more bug''. In other words, once we have found a bug, we typically run the program again to find that it is still wrong -- just in a different way. Thus, coming up with ``correct'' models is an iterative process in which the steps we discuss below will simply be repeated as often as necessary.

\subsection{General considerations about finding issues in computational models}

\subsubsection{Make it simple!}
\label{sec:simple}
As humans, we have a tendency to believe in the correctness of our work. Therefore, we tend to plow forward with implementing large parts of models before we start to assess their correctness. But this leaves us vulnerable to then having too many places where a problem may be lurking -- everything we did since we last checked that the model was correct. As a consequence, the most important piece of advice towards finding problems is to test and check frequently: If the goal is to implement a time-dependent elastoviscoplastic model on complex geometries, start with a static linear-elastic problem in a cube with simple boundary conditions; then check the correctness of the model; add the viscosity (or the plasticity); check the correctness of the resulting model; and so on. Testing every small step is a much better strategy towards building complex models than writing the entire model first and only then starting the debugging process. (In Section~\ref{sec:keeping-it-solved}, we will also outline a few strategies for making sure that parts of the code that have already been checked remain correct despite continued development of a computational model.)

A corollary to the above observation is that it is very difficult to debug complex models. Rather, if the output of that time-dependent elastoviscoplastic model on a complex geometry looks wrong, simplify it to a simple geometry, with infinite viscosity and infinite yield stress (or, better, remove these terms from the implementation). If the result is still wrong, remove the body force, simplify the boundary conditions, etc. The goal is to come up with as simple a test case as possible that illustrates the problem. Having a simple model also helps with being able to say without ambiguity that the output is wrong. For complex models, we often have a hunch that something does not look quite right, but it might be hard to pinpoint what that is; on a box geometry we can often visually say that the boundary conditions are different than what we intended to prescribe, or that the displacement points in the opposite direction of the body force we thought we provided, and this can offer important clues as to the source of the problem. Although the proposition of performing model simplification seems at first thought like a time sink, it often results in time saving when compared to the ``shoot in the dark'' approach to fixing the complex model.

\subsubsection{Build on the work of others}
\label{sec:tools}
The most consequent extension of trying to keep things simple is to not actually implement it yourself, but to build codes on the work of others. For example, nobody should be writing their own iterative or direct solvers for linear systems any more -- there are excellent software packages, developed for many years by experts in the area, that have all of this functionality, are portable to many different platforms, and are optimised to run on problem and machine sizes far beyond what most of us can access on a regular basis~\citep[e.g.][]{trilinos-web-page,heroux2005trilinos,petsc-user-ref,petsc-web-page,davis2004umfpack,amestoy2000}. The same can be said for libraries that provide everything one might ever need for the finite element discretisation ~\citep[e.g.][]{dealII94,FEBio,FenicsxBasix,Kratos,MFEM,NGSolve,libMesh}.
These packages have extensive test suites and, while they do have bugs, one can generally assume that whatever they do is vastly more likely to be correct than any code one could implement oneself. Building on others' work therefore saves enormous amounts of time on debugging, in addition of course to not having to write corresponding functionality to begin with. 

In practice, it is not uncommon that a finite element solver for a complex problem, written from scratch, would require 10,000s or even 100,000s of lines of code; when built on state-of-the-art discretisation and solver libraries, it might require one tenth or even less that amount. Empirical evidence shows that the time taken to comprehend and incorporate third-party libraries to tackle specialised tasks pays dividends surprisingly quickly when one evaluates problems of appreciable size and/or complexity.

Finally, one could even reuse complete computational solid mechanics models developed, implemented and validated by other researchers. Most of us will be happy that others expand on our models -- this is precisely why we make our full codes freely available. Caution is warranted in such cases, though, because one must dedicate time and effort to fully understanding the underpinnings of the code in order to understand whether an existing code is suitable as the basis for one's own application.

\subsubsection{Employ widely used tools for development and debugging}

An extension of the previous section is to build software using widely used (and typically free!) tools to make development and debugging more efficient. In particular, this includes integrated development environments (IDEs) such as Eclipse~\citep{eclipse}, Microsoft Visual Studio Code~\citep{visual-studio}, Apple's Xcode~\citep{xcode}, Qt Creator~\citep{qt-creator}, or equivalent tools available for nearly every programming language. IDEs are not just fancy text editors: They actually \textit{understand} the software being developed, know variable names, show tooltips that provide information about function arguments and the function's documentation, can refactor code, automatically indent code uniformly, and just make programming quicker and less error-prone.

IDEs also integrate well with debuggers. In contrast to ``debugging with \texttt{printf}'', debuggers allow executing programs one step at a time, and in the process inspecting the values of variables. Using a debugger does not only remove the edit-compile-execute cycle that slows down more manual approaches, it also provides actual \textit{insight} into a program's working -- because one can observe the program \textit{working} on its data as one steps through the program one line or function call at a time! For compiled languages, GDB~\citep{gdb} and LLDB~\citep{lldb} have long been the workhorse debuggers, and both are nicely integrated into the user interfaces of IDEs. That said, similar tools also exist for other programming languages.

Finally, there are many other widely used tools for different aspects of development and debugging. For example, Doxygen~\citep{doxygen} is a widely used tool for the generation of easily searchable documentation. Profilers and analysis tools such as Valgrind~\citep{valgrind} and its supporting tools, Heaptrack~\citep{heaptrack}, Intel VTune~\citep{vtune}, LIKWID~\citep{likwid}, and Gprof~\citep{gprof} help find performance or memory problems.

\subsubsection{Look at the solution}
A specific tool that we have almost always found useful in developing computational models is to use visualisation software to graphically represent the model's output.%
\footnote{There are many software packages that can be used to visualise and interrogate a solution. 
The two most widely used codes today are Visit~\citep{VisIt} and Paraview~\citep{ParaView}, both of which provide ways to visualise scalar and vector-valued, 1d/2d/3d solutions. They also provide tools to compute and visualise quantities derived from these solutions. Both software can be run on small laptops, but can also be used on solutions computed on thousands of processes with billions of unknowns.}
That is because, even in cases where the exact solution is unknown as a function of $x,y,z$, we often know certain things the solution must satisfy. For example, if the body's geometry, the boundary conditions, and the body forces are all symmetric with regard to a plane or point, then we know that the solution should also be symmetric -- and if it is not, we know that it is ``wrong'' even though we may not know what the ``correct'' solution is. One can generalise this approach by asking about other ``invariants'' the solution has to satisfy and that we can check even if we do not know the exact solution. For example, if a time-dependent model is incompressible, we can compute the volume of the deformed object in each time step and verify that it remains constant (to within reasonable limits relating to the numerical scheme). Likewise, if a model lacks dissipation, then the total energy needs to remain constant. In practice, with enough thought, we can often come up with \textit{many} such invariants that when checked can help build confidence that a solution is correct and can be trusted -- or, conversely, to say unambiguously that it can not be correct.

Recognise, though, that there are many cases in the world of nonlinear solid mechanics where a feasible-looking solution does not provide sufficient information to confirm that it is indeed correct in all respects. For instance, visual information is insufficient to confirm that energy and angular momentum are conserved on a global scale in dynamic problems. In problems involving large displacement increments, as is seen in problems involving snap-through behaviour or other elastic instabilities, interesting phenomena may occur in the time between time steps. This may lead one to think that the observed behaviour is wrong, which is not necessarily the case; rather, the applied numerical schemes or parameters are insufficient to capture these interesting effects.

\subsubsection{Create known solutions}
\label{sec:known-solutions}
Once we have checked that all suspected invariants are respected by our solution, it is time to compare against an exact solution. The issue is that for most complex and coupled problems, there are no known analytical solutions. But it turns out that with the right trick, they are easy to create -- a technique called the Method of Manufactured Solutions (MMS). Rather than describe this method in detail, let us refer to \citet{Roache2019,Salari2000,Jelinek2007,MFIX}. 
In the end, the method provides us with an exact solution against which we can check our numerical solution for closeness and convergence.
A second, and obvious, alternative to the ``gold standard'' MMS would be to replicate results already published (and hopefully verified and widely agreed upon) in the scientific literature.

\subsubsection{Talk to someone}

There are times when we are just unable to find what the problem is. In such cases of being truly stuck, it has often turned out to be surprisingly helpful to simply explain the issue to a colleague. Every experienced programmer can tell stories of running down the list of symptoms with a colleague, or showing them what the code does, just to stop mid-track with the words ``never mind, I know what the problem is''. What happens in such situations is that the simple process of explaining something illuminates misunderstandings, or forces critical thought about issues previously considered ``obvious''.

\begin{figure*}
	\centering{
		\includegraphics[width=0.85\textwidth]{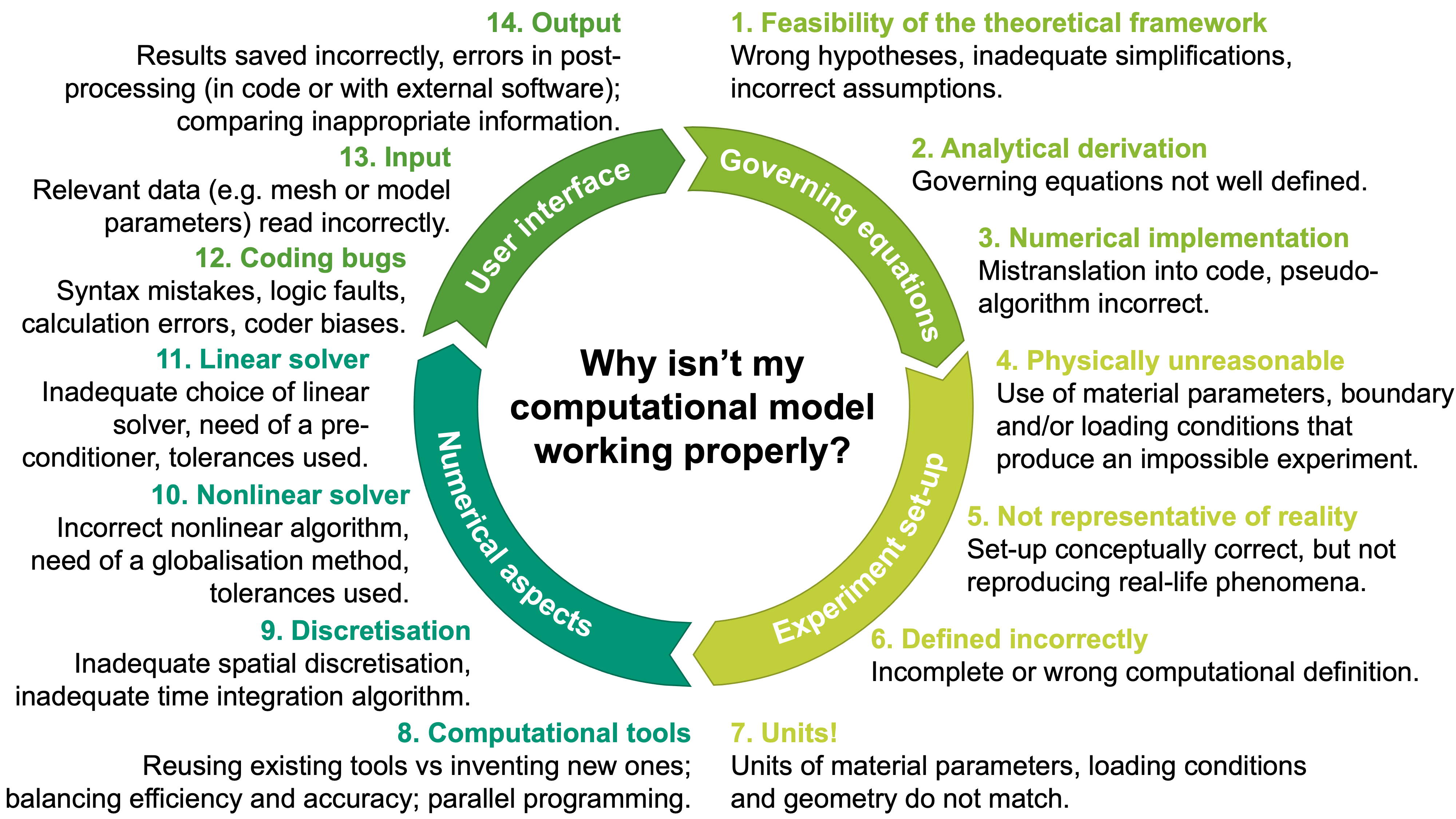}
		}
	  \caption{\it An overview of the different components of a computational model, along with possible sources of errors.}
	  \label{fig:categories}
\end{figure*}

\subsection{Things that can (and do) go wrong, and how to solve them} 
\label{sec:things-that-can-go-wrong}

Let us then move to exploring what specifically can go wrong. Fig.~\ref{fig:categories} provides an overview of the components of a computational model (shown in the central circle), along with possible mistakes one can make in each of these components (listed into categories, including a short description in black for each).

\begin{figure*}
	\centering{
		\includegraphics[width=\textwidth]{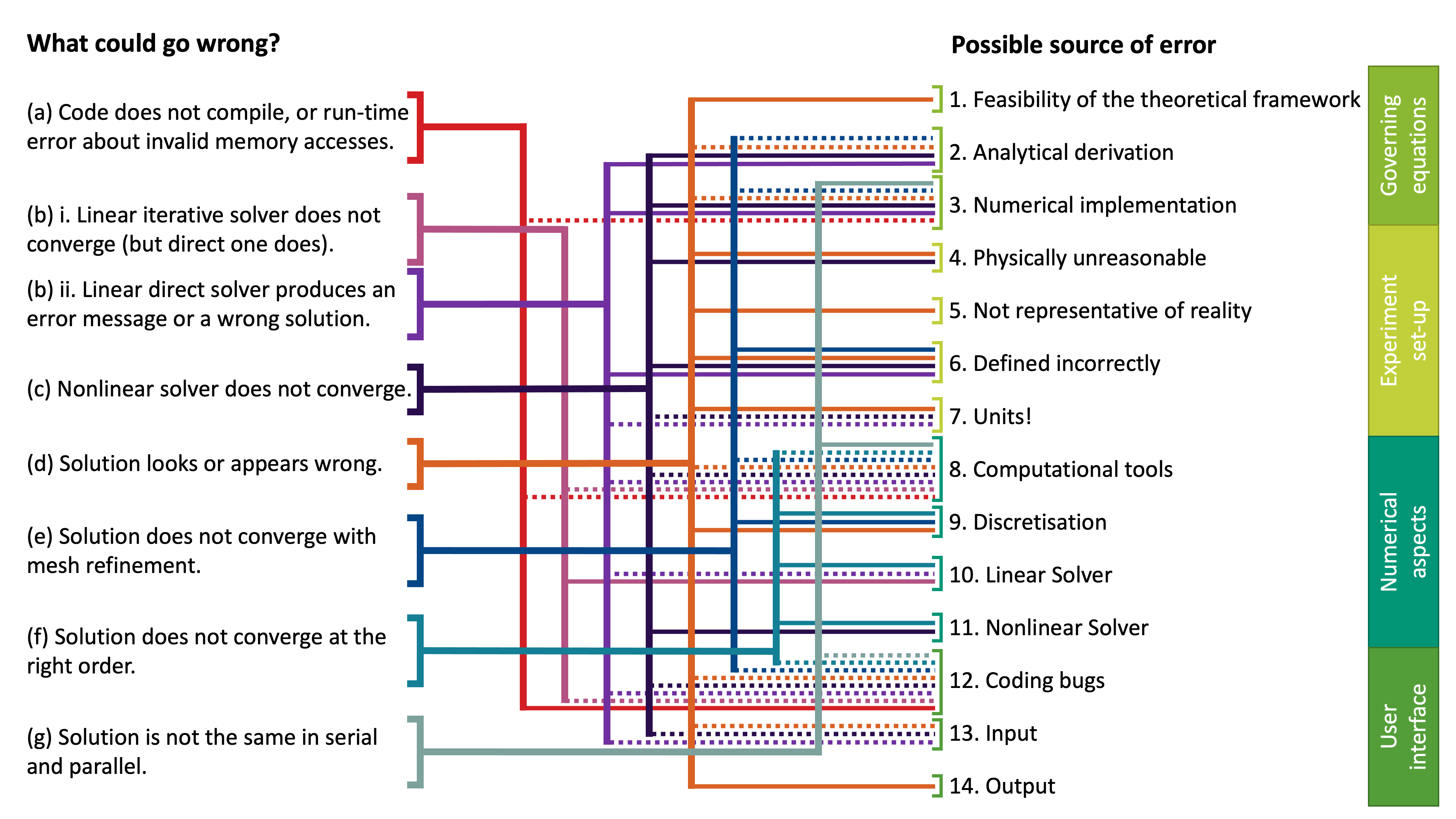}
		}
	  \caption{A schematic linking each potential problem (left, corresponding to the sub-sections of Section~\ref{sec:things-that-can-go-wrong}) to the possible sources of error (right, described in more detail in Fig.~\ref{fig:categories}). Typical root causes of the problem are indicated with solid lines, while dashed lines represent not-so-common but also possible causes.}\label{fig:error-source}
\end{figure*}

We could -- in the spirit of finding the needle in the haystack -- simply go through the entire list and question the correctness of our model in each category. That said, from a practical perspective, it is often easier to start from an empirical observation of the \textit{symptoms} of what is happening, and from there going to which of the components of a model may be wrong. 
As a consequence, in the following let us instead enumerate common symptoms of ``wrongness'', and for each discuss what that might imply for the origin of the problem.
To complement this analysis, Fig.~\ref{fig:error-source} provides a schematic of the categories of error sources (as defined in Fig.~\ref{fig:categories}) that are typically the root causes of each of these symptoms.

\paragraph{(a) If the code does not compile, or if one receives a run-time error about invalid memory accesses.}
As mentioned in Section~\ref{sec:intro}, we do not want to dwell on these errors herein. A useful starting point is to carefully read the error message by the compiler, the linker, or the run-time system of the programming language used; for example, some programming languages output concise error messages when accessing out-of-bound array elements whereas others may simply produce a segmentation fault.

\paragraph{(b) If the linear iterative solver does not converge.}
Nearly every approach to solving computational models ultimately results in the need to solve one or a sequence of linear systems, often very large but sparse ones. This can either be done using (sparse) direct solvers that use variations of Gaussian elimination to find the solution of the linear system \citep{davis2004umfpack,amestoy2000}; or iterative methods such as Conjugate Gradients (CG), Generalised Minimal Residuals (GMRES), or any number of other Krylov subspace methods \citep{saad2003iterative,templates}.
Iterative solvers sometimes do not converge, that is, they do not find the solution of a linear system even though we allow them to run for sufficiently many iterations (say, a few hundred or a few thousand iterations).

Direct solvers are rarely implemented in user code; instead, one typically uses pre-packaged solvers written by others, and so their answer can generally be considered correct. As a result, if an iterative solver does not converge, it is often useful to use a direct solver instead. If the direct solver also produces an error message, or if it produces a solution that is not correct, then the problem lies in the linear system we have given to the solver, and we need to search for mistakes in the code that builds the system matrix and right hand side, as well as in the ideas that resulted in this code.
A common source of error is the ill-specification of Dirichlet boundary conditions, leading to a singular system.
Another common problem is choosing a quadrature formula with too few quadrature points for the given problem and finite element polynomial degree, again leading to a singular linear system.

If a direct solver results in the correct answer, but an iterative solver does not, then the linear system has been correctly assembled but is solved incorrectly. Assuming that the implementation of the solver is correct, the problem must then lie in the choice of the solver itself or, as is often the case, in the choice of the preconditioner used to make the problem better behaved. For example,  the Conjugate Gradient method can only deal with symmetric and positive definite matrices, using symmetric and positive definite preconditioners. It will typically not converge if either the matrix or the preconditioner are non-symmetric or indefinite. If matrix and preconditioner are \textit{expected} to have these properties, then non-convergence should trigger a search of the code that assembles them; if they are not expected to have these properties, one needs to switch to different iterative methods such as BiCGStab, Minres, or GMRES. 
Helpful references for those unfamiliar with the matter are \citet{templates} and \citet{saad2003iterative}, which provide guidelines for the choice of an iterative solver.

\paragraph{(c) If the nonlinear solver does not converge.}
Around the linear solver sits the nonlinear solver loop: A Newton method, a fixed point (Picard) iteration, or some variation thereof. The fact that the linear solver works means that we can solve one nonlinear iteration, but if the outer iteration does not converge (that is, if the norm of the nonlinear residual vector does not decrease), then that often means that the inner solver is solving the wrong problem. 

Typically, this is because the system matrix and the right hand side do not match -- for example, in a Newton iteration, the matrix is not an algorithmically consistent linearisation of the residual (which forms the right hand side vector). These cases are awkward to debug because, for complex materials, the Newton matrix often contains a lot of not-so-nice terms; assessing the correctness of the bilinear form that leads to this matrix frequently takes pages of hand-written derivatives, and comparing them to what is implemented. A better approach, however, is to recognise that the Newton matrix is the derivative of the residual vector, which can often itself be expressed as the derivative of an energy functional \citep{Zienkiewicz2005a,Miehe2011b,Miehe2011c,Mielke2011a}. Humans should not have to implement this: Taking derivatives is something that can be left to computers, and automatic \citep{Griewank1996a,Fike2011a,Phipps2012a} or symbolic \citep{Bauer2002a,Certik2013a} differentiation libraries are happy to do this work for us. While this slows down computations, it avoids a common source of bugs. If one has verified the correctness of an implementation, one can later replace computer-assisted differentiation with hand-written code if performance is a concern -- but at least at that point, one has a base-line that is known to be correct.
This approach is particularly useful for highly nonlinear and saddle-point problems (e.g. those arising from non-convex energy functions), which might naturally include some instabilities that are indistinguishable from linearisation errors.

There are other possibilities for why nonlinear solvers may not converge. The most common one is using a Newton method and taking full steps. Newton's method is known to converge only when started close to the solution \citep{Kelley_2003}. 
If that cannot be guaranteed, one needs to use line search or another globalisation strategy to ensure that the method converges \citep{NW2006}. 
Despite their conceptual simplicity, globalisation methods are often awkward to implement, especially if taking the smallest number of nonlinear steps is a concern for performance reasons. It may be easier to rely on external libraries that provide these, such as the KINSOL package of SUNDIALS \citep{hindmarsh2005sundials}, the NOX package of Trilinos \citep{heroux2005trilinos,trilinos-web-page}, or the SNES solvers in PETSc \citep{petsc-user-ref,petsc-web-page}.

Finally, if the inner linear solver is an iterative method, its tolerance might be too loose for the solution of the linearised problem to be a useful direction for the nonlinear outer loop, and reducing the tolerance might help. This is specifically the case if the outer iteration solves for the solution (say, a Picard iteration), rather than for updates (as typically done with Newton or defect-correct iterations).

\paragraph{(d) If the solution looks or appears wrong.}
Let us assume that we have gotten to a point where our code runs without error messages, and outputs a solution that we can visualise. With a bit of physical insight, we can often tell whether it looks reasonable or not. For example, we often know that the solution should be left-right symmetric, or that given the material parameters and magnitude of forces we expect that the displacements should be on the order of a few millimetres. We expect that anisotropic materials will behave differently depending on the orientation of the loading with respect to the preferred material direction. We expect that elastoplastic and viscoelastic media will demonstrate a strong sensitivity to the load magnitude or rate. If the solution as visualised violates some of these intuitive checks, it must be wrong. 
(On the other hand, we can and should not conclude that just because the solution looks reasonable, it is actually correct. This must be verified, see below. Furthermore, one should be mindful of the limitations of the visualisation method itself -- for example, does one expect to see the intuitive continuity of the solution when hanging nodes are present or when a discontinuous Galerkin method is applied, and does my visualisation framework understand higher-order representations of finite element fields?)

The first step to finding where the problem lies is often to carefully observe \textit{in which way exactly the solution is wrong}. For example, if we have provided displacement boundary conditions, does the visualised solution satisfy the prescribed displacements? If we have prescribed a downward-pointing body force (e.g. gravity), does the body sag as expected? We have often found it useful to just write down a list of things we expect to see, and then to go down this list looking at a visualisation, either confirming or falsifying each of our predictions. Spending more time in coming up with more entries of the list of things we expect to see is often useful in narrowing down the problem. A similar step can and should be done with other invariants that can be checked in non-visual ways. Sometimes a parameter sensitivity analysis -- say, determining how the displacement changes as one increases the size of a body force or boundary traction -- aids in this step, as it assesses the dominant influences on the rendered solution and could help determine if a confluence of parameters might be problematic.

There remains the question of what to do if an invariant is not satisfied. Observation often helps: If the boundary displacement looks wrong, the boundary conditions may have been implemented wrongly. If the object is rotated or translated when it should not have been, it may be that we have not removed the nullspace of the differential operator. (This typically leads to a singular matrix with zero eigenvalues, but both direct and iterative solvers frequently succeed in finding a solution of the problem despite this issue.)
If the volume is not preserved in an incompressible model, then the incompressibility constraint is probably implemented wrongly. Seeing patterns in which invariants are or are not satisfied also helps narrow down the places where something could be wrong. 

That said, it is also possible that the original model formulation may have had mistakes, and in that case one might have to go back many steps in the loop in Fig.~\ref{fig:categories} -- certainly a daunting though fortunately not too common case. 
When the issue seems to be in the definition of a (new) constitutive equation, it is often helpful to conscientiously review the physical principles (and assumptions) behind the material model. In this way, one can challenge the choice of internal variables selected (not only the variable in itself, but also its form -- should it be a relative value instead of an absolute one, or a rate of change, or something else?). 
Even the general framework used to develop the material model might not be the most adequate.
The approach in the literature that has the most momentum might still have room for improvement. As an example, the volumetric-isochoric split might not be physically adequate to model the fibre component of anisotropic fibre-reinforced composite materials, even if it is a standard approach in the field~\citep{Sansour2008}. 
If the problem with the material model persists after examining its theoretical foundation, the bug may be in the implementation itself. In line with Sections~\ref{sec:tools} and~\ref{sec:known-solutions}, one could try using an alternative (simpler) version of the same constitutive model to pinpoint the origin of the problem. For instance, replacing a principal-stretch based formulation for an equivalent one based on strains can help isolate a problem with the eigenvector calculation.
For mixed methods (or coupled problems in general), additional points to consider include whether the residual has been expressed correctly, whether the discretisation satisfies the \mbox{Ladyzhenskaya–Babu\v{s}ka–Brezzi} (LBB) conditions \citep{Hughes2000a}, and whether the history-dependent variables are evolving in an appropriate manner. One should pay particular attention to the residual as it is this quantity that we seek to reduce, with zero as an indicator of the equilibrium solution -- an incorrect residual defines an incorrect equilibrium point.

In addition to the physics of the problem, one should also consider the role of the numerical algorithms on the computed solution. The numerical integration (quadrature) scheme and order is typically chosen such as to integrate a mass matrix exactly. An inappropriate selection of the integration order for the discretised differential equation or the polynomial order for the finite element basis functions may render incorrect results; a well-known manifestation of this would be volumetric locking in near-incompressible media or shear locking in thin, bending dominated structures. The time step size (or, for quasi-static problems, load step size) should also be chosen appropriately when the increment in the applied load is large, particularly in the case of highly nonlinear or rate- or history-dependent materials.
An uncommon but still conceivable issue with numerical algorithms is that they assume ``perfect'' conditions, which might not be mirrored in the ``real-life'' equivalent we are trying to reproduce.
To illustrate this, consider a thin cylindrical tube subjected to a compressive load. The tube will buckle and the folding pattern may be predicted to be perfectly axisymmetric.
In reality, there are both geometric and material imperfections that will break this symmetry. In such cases, we can either implement a numerical solution to the numerical problem (e.g. perturb the load), or avoid the issue altogether through changes in the modelling conditions to better reflect the ``real-life'' conditions (e.g. introduce minor deviations in the geometry and/or material characteristics).

In general, one must always be aware that simplifying assumptions (be they those made consciously, or those that are implicitly applied through the choices made in the formulation and implementation stages) might have unintended consequences, and should therefore be reviewed with more than a hint of scepticism. If they are not thoroughly analysed before implementation, then not too infrequently are limitations of the formulation and/or numerical framework the root cause of incorrect solutions. The possibility that these might need further assessment should also not be dismissed too easily, as each method undoubtedly has some drawbacks or consequences that need to be factored in. On the rare occasion, the source of error might even be traced back to the theoretical foundation upon which the method of assessing the correctness of the solution is built. No generalities can be made here, but as a concrete example two of the authors of this paper had to track down the reason for a deficiency of dissipation in a poro-viscoelastic model, only to find that it was transferred to a fundamental, but secondary, dissipation term that had been neglected.

\paragraph{(e) If the solution does not converge with mesh refinement.}
Once we have satisfied ourselves that the solution at least looks reasonable, it is time to verify that it actually is. This is best done by using a known solution, either because we have a simple-enough test case for which the solution can be derived analytically, or using the Method of Manufactured Solutions \citep{Roache2019,Salari2000,Jelinek2007,MFIX}. 
If we know the exact solution (which we will denote by \mbox{$u=u(\mathbf x)$} or \mbox{$u=u(\mathbf x,t)$} though concrete applications may of course use different symbols), we can compute the error in the numerical approximation $u_h$ through a norm such as the $L_2$ norm, \mbox{$\|u-u_h\|_{L_2} = \left[\int |u(\mathbf x)-u_h(\mathbf x)|^2 \, \text{d}x \right]^{1/2}$}.

A correctly chosen and implemented numerical scheme should of course yield numerical approximations $u_h$ that converge towards $u$ as the mesh is refined (and, if time dependent, as the time step size is reduced). In other words, we want that \mbox{$u_h\rightarrow u$}, or equivalently that \mbox{$\|u-u_h\|\rightarrow 0$}. If that is not the case, then either the computed or the exact solution is wrong.

Such cases are fortunately rare if the solution has passed the tests of the previous sub-section. If it does happen, it is often useful to output and visualise the error, \mbox{$e=u-u_h$}. Doing so then reveals how the exact and computed solution differ: Is the error large at the boundary? Then the boundary conditions are probably wrongly implemented. Does the error show a checkerboard mode? Then the choice of finite element may be questionable. If there is no pattern to the error, it may be that you are using the Method of Manufactured Solutions (see above) and have made mistakes in deriving the (often complicated) right hand side or boundary value functions -- in other words, you are solving the problem correctly, just for the wrong right hand side. As before, it is often useful to let some symbolic algebra program \citep[e.g.][]{Bauer2002a,SymPy,maple,Mathematica} compute these right hand side functions, rather than doing it by hand on a piece of paper.

\paragraph{(f) If the solution does not converge at the right order.}
Having established that the solution converges, the last remaining question is whether it does so at the correct order. In many -- though not all -- cases using the finite element method, for example, a numerical approximation $u_h$ computed using piecewise polynomials of degree $p$ will yield a solution in which the error decays like \mbox{$\|u-u_h\|_{L_2} \le C h^{p+1}$}: reducing the mesh size $h$ by a factor of two (e.g. by uniformly refining each cell of the triangulation) reduces the error by a factor of $2^{p+1}$, at least asymptotically as we keep refining the mesh.

If this is not the case, then we have either chosen an inappropriate discretisation, or the discretisation has not been correctly implemented. The former is a mathematical question for which we cannot give general guidance (at least for complex, coupled systems); the latter can often be avoided by not writing computational codes from scratch but by building on discretisation libraries such as the ones mentioned in Section~\ref{sec:tools}.

Another possible reason for lack of convergence at the right order is if one uses an iterative method for the solution of linear systems, but the tolerance with which these systems are solved (i.e. at which the method terminates iterations) is chosen too large. In such cases, the overall error is dominated by the linear solver error rather than the discretisation error, and reducing the tolerance results in recovery of the correct error order. The same can obviously happen if a nonlinear system is not solved to sufficiently small residuals.

\paragraph{(g) If the solution is not the same in serial and parallel.}
Debugging parallel programs is an art in itself, and many numerical libraries include algorithms that make parallel programming easy, safe, and deterministic.
For instance, they might incorporate frameworks that help to write distributed programs and methods to synchronise data between parallel processes, and often leverage linear solvers that work in a parallel environment.
We will therefore assume that the reader is using such a framework and is not writing raw parallel processing code themselves.

With that in mind, when augmenting a serial program to run in parallel, one primarily needs to ask oneself if the required computations are being done on the right process, and if the correct data is being transferred to other processes at the correct time.
In typical finite element programs, the assembly process can mostly be performed with each cell's work being done completely independently of another until such time that locally assembled contributions are distributed to a global matrix.
If the distribution of the linear system is not synchronised correctly then the parallel linear solver will have an inconsistent view of the global matrix on each process.
Post-solve, the distributed solution vector needs to be correctly communicated to each process such that the solution $\mathbf{u}_{h}(\mathbf{x}) = \sum_{i} \mathbf{N}_{i} (\mathbf{x}) u_{i}$ on each (local) finite element can be correctly reconstructed using the correct solution coefficients (or degree-of-freedom values) $u_{i}$ and the vectorial basis functions $\mathbf{N}_{i}$.
Failure to do so might result in visualisation artefacts in the best case, or divergence of the numerical method in the worst case.

As before, if the solver produces a solution, careful visual inspection often helps understand where a problem may be. If, for example, there are artefacts at the boundaries of subdomains owned by different processors, there is likely a problem in pre-solve assembly -- or maybe the post-processing routines also need to be adapted for the parallel setting to ensure that every processor knows that part of the distributed data necessary to create visualization files. A general rule in debugging parallel programs is to also follow Section~\ref{sec:simple}: Make it simple, for example by testing whether a program that produces wrong results with 100 million unknowns on 256 processors also produces wrong results with 200,000 unknowns on 2 processors. The latter will generally be much easier to debug.

\section{Keeping problems solved}
\label{sec:keeping-it-solved}
Rarely do we develop a computational model, debug it, apply it to a concrete situation, appreciate the fact that its predictions match physical measurements, and then put the model onto a shelf (or switch the file permissions to ``read-only'' as it may be). Rather, a successful model typically serves as the starting point for another model in which we change some of the physical conditions that describe a situation or modify the material's constitutive relations. 

In practice, making these modifications will then lead to a model that, in all likelihood, will again be wrong on first attempt. To debug it, we could again assume that \textit{everything is suspect} as mentioned in Section~\ref{sec:what-is-wrong}. But we have built on something that worked before, could we not simply assume that the problem must be in what is new? The answer is \textit{not generally}: In modifying the previous model, we probably changed parts of the code (or the formulation) and thereby may have broken the previous model. This is unsatisfying because it means that we cannot trust any part of the model. On the other hand, there are some strategies that allow us to deal with the situation in a more efficient way, and we will discuss these in the current section.

\subsection{Incremental development: Test frequently, use version control}

In keeping with the recommendations of Section~\ref{sec:simple}, the best way to avoid getting into a situation where \textit{everything} is suspect is to make incremental changes, check that the output of a previous test is unchanged and still correct, and then commit the current state of the project to a version control system such as Git \citep{git,chacon2014pro}. Using version control has many advantages. For the current purpose, it includes being able to exactly see what has changed since the last commit (and consequently the last time the model was checked), vastly reducing the places where one might have to look for newly introduced bugs. It also allows rolling back to the last committed state where we knew that the solution was correct if we really cannot find what the problem is -- and then starting the development of the current feature from scratch.

Another advantage of version control is that we can be unafraid of drastic changes. For example, as explained in Section~\ref{sec:simple}, it is difficult to debug complex models. If we suspect that the implementation of the boundary conditions are the problem, would it not be nice to just remove the body force, the nonlinear loop, the coupling to other variables, the time loop, replace the complex mesh by a much simpler one, and just strip the code to its bare minimum that could illustrate just the handling of boundary conditions? If a code is not under version control, this is a slightly scary proposition because we have to rely on our own diligence in keeping track of the state of the code. With version control, we can just hack away at the code, find the bug in the minimised version, and try a fix; if it works, we just save the few lines of changes, tell the version control system to revert to the last committed state, and re-apply the few changes we found to fix the problem. 

Using version control systems has been found so useful that there is really no excuse any more today to not use one -- switch yesterday! It will be time well spent, and will afford you the freedom to experiment without the fear of permanent consequences.

\subsubsection{Use a test suite}
Incremental development as suggested above is only easy if checking the continuing correctness of a code is easy. It is not if checking involves tedious manual work, changes to the code (for example commenting in the right hand side of a manufactured solution, and commenting out the ``real'' one), and comparison by hand/eye whether the solution continues to be correct.

A better design uses automated checks. For whole applications, a common approach is to drive the code through external input files that describe right hand sides, boundary conditions, formulations, and what output quantities should be computed. Automated testing frameworks can then compare the results of a simulation with a given input file against known-to-be-correct answers for this input file. Incremental development must then be done in a way so that past input files continue to be valid, and new development implements features that are then tested by new input files that are added to the test suite. While this may seem like a lot of work, it is not in practice -- as shown, using concrete examples, in \citet{THB15}.

Test suites can be written from scratch but in the spirit of Section~\ref{sec:tools}, we would be remiss if we advocated for this approach. Instead, let us refer to gtest~\citep{gtest}, ctest~\citep{ctest}, and catch2~\citep{catch2}: all widely used libraries that make many aspects of automatic and frequent testing much simpler.

\subsubsection{Incorporate benchmark problems into tests}

As useful as it is to develop a test suite that runs checks on one's own metrics, evaluating the output of programs against benchmarks from the established literature really helps to solidify that one's work is correct and remains so.
There are an abundance of simple and well understood benchmark problems that have been studied to the ends of the earth, and it is natural (and good practise) to incorporate at least some of them into a test suite for a scientific code. Many relevant papers in each field about numerical methods use such benchmarks, and replicating these in your own work is a good starting point.

\subsubsection{Combine testing with refactoring}

Incremental development, the use of version control, frequent testing, and the use of an automated test suite combine well with one of the most useful techniques for working on existing codes: Refactoring -- that is, the continuous reshaping of a code base to make it fit for new features, to make it easier to find and fix bugs, or to just improve the overall quality of code. An excellent description of refactoring and concrete refactoring steps can be found in~\citet{Fowler}. Refactoring only works well if one has a sufficiently broad test suite so that one can be sure that a small change to the code base is correct if all tests in the test suite succeed, and if that test suite can be run frequently and automatically.

As discussed in \citet{Fowler}, refactoring code should also be used as an opportunity to extend the test suite. A common way to do this is to write ``unit tests'' as a first step of a refactoring session.

\subsection{Define and enforce a quality standard}
By defining a quality standard that we would want to adhere to, we are essentially writing a contract for ourselves to prevent ourselves from employing poor practices, poor judgement, and minimising the oversight of issues that might arise during the development process.
Although this might seem somewhat obvious, we have probably all been guilty of cutting corners somewhere during the scientific process, perhaps to the detriment of the current and future work's outcome.
For example, honest thought will suggest that rushing to implement something to a timeline without checking its correctness (or even worse: knowingly ignoring errors emitted when a code is tested in debug mode) is likely going to lead to more work downstream in the best of cases -- and to wrong results in a publication in the worst of cases.
Having a semi-rigorous approach to viewing work quality would navigate one away from such scenarios.

One way to deal with the pressures of development is to map out and plan what one envisions to be the remainder of the project, early on in each project's life.
In doing so, a rough timeline for the work can be established and from that one would be able to identify and set a pragmatic set of quality targets that one strives to maintain.
These could be related to the implementation (e.g., adding checks as debugging aids, improving code quality and reducing redundancy, only accepting code that has been independently tested or verified), adding unit tests (e.g., aimed at validating constitutive laws, finite element formulations, etc.), implementing new technologies to assist in the development process (e.g., switching to version control, using continuous integration tools), or simply even actively learning new skills, or improving existing ones, to increase the quality of the next piece of work that is to be done.
Planning also helps prevent one from repeating old mistakes, as one could clearly identify an upcoming pitfall and apply remediation strategies before work has begun.
Hitting quality targets would naturally maintain the momentum of improving existing (and future) code quality while considering the following interesting feature to tackle next.

In the end, holding oneself to high quality standards is also an important part of our own professional ethics. For example, the Code of Ethics for Engineers by the National Society of Professional Engineers states both ``Engineers shall not complete, sign, or seal plans and/or specifications that are not in conformity with applicable engineering standards. If the client or employer insists on such unprofessional conduct, they shall notify the proper authorities and withdraw from further service on the project.'' (Section III.2.b) and ``Engineers shall accept personal responsibility for their professional activities'' (Section III.8., \citet{nspe-ethics}). Having pride in one's own work also requires being able to honestly answer ``yes'' when asked if one is \textit{sure} that a computed solution is correct. As a consequence, if a code cannot be completed and sufficiently tested by a deadline, then results must clearly be marked as preliminary, or the paper or proposal needs to be delayed to the next deadline.

\section{Conclusions}
\label{sec:conclusions}
Developing software for the purpose of computational engineering inevitably means spending a lot of time and effort determining the source of issues with mathematical formulations, as well as their numerical implementation in the form of generic computer code (a framework) and specific simulation configurations.
Debugging any of these aspects is a difficult process, as is being able to find a foothold from which to establish the source of non-programmatic errors.
In the end, these are both skills that need to be learned and practised; having a companion who has gone through this process many times over provide some insights into where to start and what to look for can greatly accelerate the process of learning.

In this paper we have provided just that: having summarised the components of the typical solid mechanics model, we have listed common categories of errors and elucidated as to why they appear in the first place.
We have then given clear -- albeit non-exhaustive -- recommendations on how to identify which category the reader's issue might be associated with and some general guidance as to how to start the process of correcting said issue.
We then conclude with some suggestions as to how the reader can ensure that their simulation framework remains reliable, and how they can improve their process of future development.

As a final comment, we wish to reaffirm the reader that developing numerical software is indeed challenging.
Being patient and allowing yourself time to work through problems of the nature presented in this paper is a crucial element of success.
The process of problem solving in this arena will get easier over time as you become exposed to more issues, be they in your own work and field of expertise, or someone else's.
With experience comes a shift in the balance of where you spend your time during development.
This will ultimately end in you having more fun and a greater opportunity to experience that unique sense of satisfaction of a successful implementation, and explore and enjoy the beauty and insights that computational physics simulations provide our curious minds!

\paragraph*{Acknowledgements.}
EC and JPP wish to thank their former supervisor Paul Steinmann for the inspiration to write this paper, which can be traced back to the talk we prepared for the ECCM-ECFD conference held in Glasgow in 2018.

EC's work was partially supported by the European Union's Horizon 2020 research and innovation program under the Marie Sk\l{}odowska-Curie grant agreement No 841047.
WB's work was partially supported by the National Science Foundation under award OAC-1835673; by award DMS-1821210; by award EAR-1925595; and by the Computational Infrastructure in Geodynamics initiative (CIG), through the National Science Foundation under Award EAR-1550901 and The University of California -- Davis.


\end{document}